\documentclass[12pt]{article}
\input epsf
\begin{document}

\title{Antisymmetrised $2p$-forms generalising
curvature $2$-forms II: a $p$-hierarchy of Reissner-Nordstrom type
metrics in dimensions $d>2p+1$}

\author{{\large A. Chakrabarti}$^{\star}$,
and {\large D. H. Tchrakian}$^{\dagger}$ \\ \\
$^{\star}${\small Centre de Physique Th\'eorique\footnote{Laboratoire
Propre du CNRS UMR7644},
Ecole Polytechnique, F-91128 Palaiseau, France}\\ \\
$^{\dagger}${\small Department of
Mathematical Physics, National University of Ireland Maynooth,} \\
{\small Maynooth, Ireland} \\ {\small and} \\
{\small School of Theoretical Physics -- DIAS, 10 Burlington Road,
Dublin 4, Ireland }}

\newcommand{\dd}{\mbox{d}}\newcommand{\tr}{\mbox{tr}}
\newcommand{\ee}{\end{equation}}
\newcommand{\be}{\begin{equation}}
\newcommand{\ii}{\mbox{i}}\newcommand{\e}{\mbox{e}}
\newcommand{\pa}{\partial}\newcommand{\Om}{\Omega}
\newcommand{\vep}{\varepsilon}
\newcommand{\bfph}{{\bf \phi}}
\newcommand{\lm}{\lambda}
\def\theequation{\arabic{equation}}
\renewcommand{\thefootnote}{\fnsymbol{footnote}}
\renewcommand{\r}[1]{(\ref{#1})}
\newcommand{\bfR}{{\sf R\hspace*{-0.9ex}\rule{0.15ex}%
{1.5ex}\hspace*{0.9ex}}}
\newcommand{\N}{{\sf N\hspace*{-1.0ex}\rule{0.15ex}%
{1.3ex}\hspace*{1.0ex}}}
\newcommand{\Q}{{\sf Q\hspace*{-1.1ex}\rule{0.15ex}%
{1.5ex}\hspace*{1.1ex}}}
\newcommand{\C}{{\sf C\hspace*{-0.9ex}\rule{0.15ex}%
{1.3ex}\hspace*{0.9ex}}}
\renewcommand{\thefootnote}{\arabic{footnote}}

\newcommand{\ie}{{\it i.e.}}

\newcommand{\ble}{\begin{array}[b]{l l}}
\newcommand{\ele}{\end{array}}
\newcommand{\blle}[1]{\begin{array}[b]{l l l}}
\newcommand{\elle}{\end{array}}

\maketitle

\begin{abstract}
The $p$-hierarchy of Schwarzschild type metrics obtained in a preceing paper is
generalised here to a corresponding Reissner-Nordstrom (RN) type hierarchy in the
presence of a point charge $q$ in $d$-dimensions. Certain special features arising,
concerning the horizons and the interior region, are discussed.
\end{abstract}
\medskip
\medskip

Generalised gravitational Lagrangians
constructed out of antisymmetrised $2p$-forms ($p>1$) were studied in previous papers
\cite{CT,OT,CT2}. For $p=1$ one obtains the usual Einstein--Hilbert Lagrangian for dimensions
$d\ge 4$, in the case with vanishing torsion. For $p>1$ a generalisation of Schwarzschild
type solutions were obtained~\cite{CT2} for $d>2p+1$. For $d=2p$, the Lagrangian
density represents a topological field theory given by the Euler density. For $d=2p+1$ one
has a generalisation of the $(d=3,p=1)$ case for higher dimensions, which will be
considered separately elsewhere. Here we give a brief presentation of a generalisation
of the Reissner--Nordstrom (RN) solutions for dimensions $d>2p+1$ and $p>1$.

The construction of antisymmetrised $2p$-forms is given in \cite{CT2}. Let us recapitulate
the essential features of these.

One starts with the tangent vector $1$-forms $e^a$, the antisymmetric Levi-Civita
spin--connections $1$-forms $\omega^{ab}$  satisfying
\[
\dd e^a +\omega^{ab}\wedge e_b =0\ ,
\]
and the curvature $2$-forms $R^{ab}$
\be
\label{1}
R^{ab}=\dd \omega^{ab}\ +\ \omega^a{}_c \wedge \omega^{cb}\ ,
\ee
where we have used the notation
\[
e^a =e_{\mu}^a \ dx^{\mu}\ ,\ \omega^{ab}=\omega_{\mu}^{ab}\ dx^{\mu}\ ,\
R^{ab}=R_{\mu \nu}^{ab}\ dx^{\mu}\ dx^{\nu}\ .
\]

Here $R^{ab}$ corresponds to $p=1$. For $p=2$ one defines
\be
\label{2}
R^{abcd}=R^{ab}\wedge R^{cd} + R^{ad}\wedge R^{bc} +R^{ac}\wedge R^{db}
\ee
and for successive increasing values of $p$ one has the recursion relation
\be
\label{3}
R^{a_1 a_2 ...a_{2p}}=R^{a_1 a_2}\wedge R^{a_3 a_4 ...a_{2p}}\ +\ {\rm cycl. \ perms.}
(a_2 ,a_3 ,...,a_{2p})
\ee
involving $3.5...(2p-3)(2p-1)$ terms of the type
\[
R^{a_1 a_2}\wedge R^{a_3 a_4}\wedge ...\wedge R^{a_{2p-1} a_{2p}}\ .
\]

This generalisation ensures that only the quadratic power of the velocity fields (namely the
derivatives of the metric or the {\it vielbeins}) appears in the Lagrangian density defined to be
\be
\label{4}
L_{(p)} = \sqrt{|g|}\ R_{(p)}\ ,
\ee
where the $p$-Ricci scalar $R_{(p)}$ is given in terms of the $p$-Ricci tensor~\cite{CT2}
\be
\label{5}
R_{(p)}{}_{b_1}^{a_1}=\sum_{(a_2 ,a_3,...,a_{2p})}\ R_{b_1 ,a_2 ,...,a_{2p}}^{a_1 ,a_2 ,...,a_{2p}}
\ee
as
\be
\label{6}
R_{(p)}=\sum_a R_{(p)}{}_a^a \ .
\ee

In the tangent frame basis ($e$-basis) the variational minima are obtained for the $p$-Ricci flat
case~\cite{CT2} from
\be
\label{7}
G_{(p)}{}^a{}_b =R_{(p)}{}^a{}_b -\frac{1}{2p} g^a{}_b R_{(p)} =0
\ee
leading to
\be
\label{7a}
R_{(p)}{}^a{}_b =0\ .
\ee

More generally, there are terms on the right hand side due to the cosmological constant
$\Lambda$ and the stress--tensor $T^a{}_b$. Such cases will be considered below.

In \cite{CT2}, though considering only static spherical symmetry, we started with the
Kerr-Schild (KS) form of the metric with the intention of an eventual passage to the
axially symmetric stationary case generalising the Kerr metric in $d$-dimensions to
$p>1$. However, certain convenient features of the diagonal--metric ansatz were
pointed out. In the present work, we employ only the latter form. Let us recapitulate it
briefly. One has
\be
\label{9}
ds^2 =\mp N\ dt^2 +N^{-1}\ dr^2 +r^2 d\Omega_{(d-2)}^2
\ee
where
\be
\label{10}
d\Omega_{(d-2)}=d\theta_1^2 +(\sin \theta_1)^2 d\theta_2^2 +...+
(\prod_{n=1}^{d-3} \sin \theta_n)^2 d\theta_{d-2}^2\ .
\ee

One sets
\be
\label{11}
e^a =\sqrt{g_{aa}}\ dx^a \ \qquad x^a =t,r,\theta_1 ,...,\theta_{d-2}\ ,
\ee
(for Lorentz signature $g_{tt}$ is replaced by $|g_{tt}|$), and
\be
\label{12}
\omega^{ab}=\frac{1}{\sqrt{g_{aa}g_{bb}}} \left( (\pa_b \sqrt{g_{aa}})e^a
-(\pa_a \sqrt{g_{bb}})e^b \right)\ .
\ee

Setting
\be
\label{13}
N=1-L\ ,
\ee
where $L(r)$ is a function of $r$ only. One obtains \cite{CT2} the very simple diagonalised
$2$-curvatures (for $p=1$, with $L'=\frac{dL}{dr}$ etc.),
\begin{eqnarray}
R^{tr}&=&\frac{1}{2} L''\ e^t \wedge e^r \ , \qquad R^{ti}=\frac{1}{2r} L'\  e^t \wedge e^i \nonumber \\
R^{ri}&=&\frac{1}{2r} L' \ e^r \wedge e^i \ , \qquad R^{ij}=\frac{1}{r^2} L\ e^i \wedge e^j \nonumber
\end{eqnarray}
with $i,j=1,...d-2$, leading to, for $p=1$, to the non-zero Ricci tensor components
\begin{eqnarray}
R_t^t &=&R_r^r =\frac{1}{2r^2} (r\frac{d}{dr})[(r\frac{d}{dr})+(d-3)]\ L \nonumber \\
R_i^i &=&\frac{1}{r^2} [(r\frac{d}{dr})+(d-3)]\ L \label{14}\ .
\end{eqnarray}
Here the indices $t,r,i$ are to be considered as frame indices like $a,b$. To avoid confusion one might
denote them as $\hat t,\hat r,\hat i$ etc. We do not use this last notation however, sufficing instead
with this warning in \r{14}.

For brevity we present directly the final results for $R_{(p)}{}_b^a$ and $R_{(p)}$, defined in \r{5}
and \r{6}, for $p>1$ and $d>2p$. These are our main results.

One obtains only the following nonvanishing components:

\begin{eqnarray}
R_{(p)}{}_t^t &=&R_{(p)}{}_r^r =(2^p p!\ r^{2p})^{-1}(d-2)(d-3)...(d-2p+1)(r\frac{d}{dr})
[(r\frac{d}{dr})+(d-2p-1)]\ L^p \nonumber \\
R_{(p)}{}_i^i &=&(2^p p!\ r^{2p})^{-1}2(d-3)...(d-2p+1)\times \nonumber \\
&&\times [(p-1)(r\frac{d}{dr})+p(d-2)]
[(r\frac{d}{dr})+(d-2p-1)]\ L^p \nonumber \\
\frac{1}{2p}R_{(p)}&=&(2^p p!\ r^{2p})^{-1}(d-2)(d-3)...(d-2p+1)\times \nonumber \\
&&\times [(r\frac{d}{dr})+(d-2p)][(r\frac{d}{dr})+(d-2p-1)]\ L^p\ . \label{15}
\end{eqnarray}

The Schwarzschild-like $p$-Ricci flat metrics of \cite{CT2} are obtained, by virtue of the last
operator $[(r\frac{d}{dr})+(d-2p-1)]$ in each member of \r{15}, by setting
\be
\label{16}
L^p\ =\ \frac{c_1}{r^{d-2p-1}}\ ,
\ee
where $c_1$ is an arbitrary constant.

If one introduces a point charge $q$ in $d$-dimensions, the corresponding stress-tensor is
given by~\cite{T,MP}
\be
\label{17}
T_t^t \ =\ T_r^r \ =\ -T_i^i \ =\ -\frac{(d-3)^2q^2}{2r^{2(d-2)}}\ , \qquad i=1,2,...,d-2\ .
\ee

Now with the inclusion of the static Maxwell field due to a point charge \r{7} is modified to read
\be
\label{18}
G_{(p)}{}^a{}_b\ =\ \kappa T^a{}_b\ ,\qquad \kappa =16\pi G\ ,
\ee
and it can easily be verified that
\be
\label{19}
L^p\  =\ \frac{c_1}{r^{d-2p-1}}\ -\ \frac{c_2}{r^{2(d-p-2)}}
\ee
generalises the RN metric in $d$-dimensions~\cite{T,MP}, for the $p>1$ systems. The condition,
$q^2>0$, for the reality of the charge implies that $c_2>0$, and
\be
\label{20}
\frac{1}{2}\ \kappa^2 \ q^2\ =\ (2^p p!\ r^{2p})^{-1}(d-2)(d-4)...(d-2p+1)(d-2p)\ c_2\ .
\ee

Thus the function $N(r)$ in \r{9} is finally given by
\be
\label{21}
N\ =\ 1\ -\ \left(\frac{c_1}{r^{(d-2p-1)}}\ -\ \frac{c_2}{r^{2(d-p-2)}} \right)^{\frac{1}{p}}\ .
\ee
For $p=1$, this reduces to
\be
\label{22}
N\ =\ 1\ -\ \frac{c_1}{r^{(d-3)}}\ +\ \frac{c_2}{r^{2(d-3)}}
\ee
which is the standard RN metric in $d$-dimensions~\cite{T,MP} with the usual notations
\be
\label{23}
c_1 \ =\ 2M\ ,\quad c_2 \ =\ D^2\ .
\ee
In \r{21} likewise, we will set  $c_1>0$ and $c_2>0$. For $p>1$ several new features arise,
which are discussed below.

For $q=0$ ($c_2=0$) one gets back the "$p$-Schwarzschild" black holes obtained in
\cite{CT2}, where their horizons, maximal (Kruskal-type) extensions and periodicity for
Euclidean signature, were studied.

For $p>1$ and $c_2>0$, there are {\em two} critical points to be considered, namely
\begin{eqnarray}
N&=&0\ \ (L=1)\ \ {\rm for}\ \ r=r_H\ \ {\rm say}\ ,\nonumber \\
N&=&1\ \ (L=0)\ \ {\rm for}\ \ r=r_0\ \ {\rm say}\ .\label{23}
\end{eqnarray}

The possibility of several {\em real} horizons will be analysed below. We will be
interested in cases where there is {\em at least one real
horizon} and hence no naked singularity at $r=0$. For each real solution one must have,
setting $L=1$ and hence $L^p=1$ in \r{21},
\be
\label{24}
c_1\ r^{(d-3)}\ -\ c_2\ =\ r_H^{(d-2p-1)}\ >\ 0\ ,
\ee
whereas setting $L=0$,
\be
\label{24a}
c_1\ r_0^{(d-3)}\ -\ c_2\ =\ 0\ .
\ee

Hence, always,
\be
\label{25}
r_0\ <\ r_H\ ,
\ee
namely that, {\em the point $r_0$ always lies inside the innermost real horizon}. Hence the
metric \r{21} can be considered for the exterior of the region $\{ r\ge r_H  \}$ without reference
to $r_0$. In the interior region however, $L$, and hence $N$, becomes in general {\em complex}
for $r<r_0$, depending precisely on the determination of the $p$-th root in \r{21}. For $r>r_0$,
we have implicitly chosen the real root in \r{21}, namely
\be
\label{26}
m=0\ \ \ \ {\rm in}\ \ \ \ e^{\frac{2im\pi}{p}}\ (L^p)^{\frac{1}{p}}\ \ \ \ {\rm for}\ \ L^p>0\ .
\ee

For odd $p$, choosing a suitably different determination for $L^p<0$, one can again have
real $N$. This is readily seen by writing \r{26}, for $L^p<0$, as
\[
e^{\frac{(2m+1)i\pi}{p}}\ (|L^p|)^{\frac{1}{p}}\ .
\]

It should be noted however that all the components of the "$p$-Riemann" tensor, from which
the "$p$-Ricci" tensors in \r{15} are obtained, depend on $L^p$. Hence they remain {\em real
even for} $r<r_0$. Moreover for $r=r_0$, they all vanish and then change sign exactly as
for the $p=1$ case.

Let us now consider the horizons. For $p=1$ \r{22} leads to the well known quadratic
equation in $r^{(d-3)}$ for the horizons. The roots can be complex conjugate, two
distinct real ones, or coincident ones (in the "extreme" RN case). For $p>1$ our equation
\r{21} leads to equations of higher order. The simplest among these is obtained for
\[
d\ =\ 4p-1\ .
\]
In this case~\footnote{It happens that these dimensions are the only ones in which the
'monopoles' of the generalised Yang-Mills--Higgs systems can saturate their Bogomol'nyi
bounds, i.e. that they are self--dual. Self--dual Yang--Mills fields on various gravitational
backgrounds are studied in \cite{BCT}.}
, the condition $N=1-L=0$ leads to $L^p=1$, and then  \r{21} yields
\be
\label{28}
r_H^{6(p-1)}\ -\ c_1\ r_H^{2(p-1)}\ +\ c_2\ =\ 0\ .
\ee

Setting
\begin{eqnarray}
x=&r_H^{2(p-1)}\ ,\ \ \ c_1=3a_1\ ,\ \ \ \ c_2=2a_2\ \ \ \ (a_1 ,a_2)>0\nonumber \\
&b^{\pm 3}=(1-\frac{a_2}{a_1^3})\ \pm \ ((1-\frac{a_2}{a_1^3})-1)^{\frac{1}{2}} \label{29}
\end{eqnarray}
the roots of \r{28}, i.e. $x^3-3a_1 x+2a_2=0$, are
\begin{eqnarray}
x_1\ &=&\ a_1(1+b+b^{-1}) \nonumber \\
x_2\ &=&\ a_1(1+be^{i\frac{2\pi}{3}}+b^{-1}e^{-i\frac{2\pi}{3}}) \nonumber \\
x_3\ &=&\ a_1(1+be^{-i\frac{2\pi}{3}}+b^{-1}e^{i\frac{2\pi}{3}}) \label{30}\ .
\end{eqnarray}

\noindent
{\bf Case 1:}  $a_2\ >\ 2a_1^3$

One has one negative root and a pair of complex conjugate ones. Hence there is no
real $r_H>0$ giving a horizon.

\noindent
{\bf Case 2:} $a_2\ <\ 2a_1^3$

One has one negative and two positive real roots and hence two distinct horizons. This case
is like the standard $p=1$ RN metric.

\noindent
{\bf Case 2:} $a_2\ =\ 2a_1^3$

The two real real positive roots coincide giving a single horizon. This case is like the standard
$p=1$, extreme RN metric.

The cubic \r{28} is the simplest possibility for $p>1$. A quartic and a quintic arise, respectively,
for the following cases.

For $d=3p$ one obtains, instead of \r{28} and \r{30}, setting $x=r_H^{p-1}$,
\be
\label{31}
x^4-c_1x^3+c_2=0\ .
\ee

For $d=3(2p-1)$, one obtains, setting $x=r_H^{2(p-1)}$,
\be
\label{32}
x^5-c_1x^3+c_2=0\ .
\ee

The extrema of these equations and of the general case $L^p=1$, can be easily located.
They lead to constraints on the existence of real positive roots (horizons). These will not
be analysed further here.

The regularisations of the "$p$-Schwarzschild" horizons presented in \cite{CT2} can
be generalised to, for example, the cases {\bf 2} asnd {\bf 3} following \r{30}. This analysis
will not be carried out here.

\medskip

\noindent
{\bf Remarks:}

We have pointed out that below a point $r_0$, always within the innermost
horizon, there is a transition to a complex metric. The components of the "$p$-Riemann"
tensor in this case are real nonetheless. If instead of considering a point charge one
considered instead a thin spherical shell as in \cite{B} (and references therein), the
interior metric can then be taken to be flat. This avoids complexification but there arise
attendant problems of gravitational instability. This possibility is not explored here.

Considering each value of $p$ separately it has been shown how the $p$-Schwarzschild
solutions of \cite{CT2} can be quite simply generalised the corresponding $p$-RN ones.
The attractive feature of this result is the simplicity of these solutions. A more realistic
approach would be consider a Lagrangian resulting from the superposition of  several
 $R_{(p)}$'s for different values of $p$, starting with $p=1$. In such a case, the solutions can be constructed numerically, and will be pursued elsewhere.

\medskip

\noindent
{\bf Acknowledgement:} We are grateful to G.M. O'Brien for useful discussions. This work was carried out in the framework of the
Enterprise-Ireland/CNRS programme, under project FR/00/018.

\small{

 }

\end{document}